\documentclass[prd,onecolumn,groupedaddress]{revtex4-1}
\usepackage{graphicx}
\usepackage{dcolumn}
\usepackage{bm}

\newcommand{\be}{\begin{eqnarray}}
 \newcommand{\ee}{\end{eqnarray}}
 \newcommand{\nee}{\nonumber\end{eqnarray}}

\begin{document}

\title{Monotone  Riemannian metrics and dynamic structure factor in condensed matter physics }

\author{ N. S. Tonchev}

\affiliation{
 Institute of Solid State Physics, Bulgarian Academy of Sciences,
1784 Sofia, Bulgaria}

\begin{abstract}
An analytical approach is developed to the problem of computation of monotone Riemannian metrics (e.g. Bogoliubov-Kubo-Mori, Bures, Chernoff, etc.) on the set of quantum states.
The obtained expressions  originate  from the Morozova, $\breve{C}$encov and Petz correspondence  of monotone metrics
 to operator monotone functions.
The used mathematical technique  provides analytical expansions in terms of the thermodynamic mean values of iterated  (nested) commutators  of a model  Hamiltonian $T$
 with the operator $S$ involved through the  control parameter $h$.  Due to the sum rules for the frequency moments of the dynamic structure factor new  presentations for the monotone Riemannian metrics  are obtained. Particularly, relations between any monotone Riemannian metric and the usual thermodynamic susceptibility or the variance of the operator $S$ are discussed.  If the symmetry
  properties of the Hamiltonian are given in terms of generators of some Lie algebra, the obtained expansions may be  evaluated in a  closed form. These issues are tested on a class of  model systems studied  in condensed matter physics.

\end{abstract}


\maketitle


\section{Introduction}
In recent decade  different concepts emerging  from quantum-information geometry \cite{BZ06,H06,PG11,DS14}  have been intensively incorporated in condensed matter physics \cite{Cr07, ZVG07,ZGC07,YLG07,ZQWS07,SMK09,QC09,S10,V10,Gu10,PL11,DL13,B14}.  The underlying idea may be briefly reviewed as follows. The properties of   macroscopic phases of matter should be encoded in the structure of the quantum state. It is defined by a density matrix $\rho$ depending  via the Hamiltonian ${\mathcal H}$ on a set of parameters, e.g. coupling constants $\lambda $ and external "fields" $ h $. If ${\mathcal H}$ smoothly depends  on the coupling constants and external fields one has a map between $(\lambda , h,$  the inverse temperature $\beta$)  and  the  set of density matrices:
$\rho\equiv\rho (\beta,\lambda, h):= [Z(\beta,\lambda,h)]^{-1}\exp[-\beta {\mathcal H}(\lambda,h )],$
where $Z(\beta,\lambda, h )$ is the partition function of the system. Further,  the set  of  density matrices $\rho$ should be endowed with a metric  structure and thus explored as a Riemannian manifold.
The realization of the set of  density matrices $\rho$  as a Riemannian manifold implies
to define a relevant distance   between two mixed quantum states \cite{BZ06,DS14}. The notion of distance  has found applications in  different fields.  For example:  thermodynamic of small systems \cite{Cr07,DL13}, geometrical description of phase transitions and quantum criticality \cite{ZVG07,ZGC07,YLG07,ZQWS07,SMK09,QC09,S10,V10,Gu10,PL11}, quantum estimation of Hamiltonian parameters \cite{BC94,P2009} and hypothesis testing and discrimination of states \cite{ACM07,CMM08,ANSV08,ZMLW14},   are  only a part of them.   If we consider  the distance  between two states  obtained by an infinitesimal change  in the values of the parameters  that specifie the quantum state, we come to the notion of a metric tensor, i.e. the set of coefficients of the linear element $ds^2$ when written  as a quadratic form in the differentials of these parameters.
  In the present study the infinitesimal variation of a single parameter, e.g. the field $h$ is considered, and the notations of a linear  element and metric will be used interchangeably.

  For our purposes we shall recall some  metrics  which are interesting  in the context of their physical applications.

The metric $d^2_{BKM}$,  after the name of  Bogoliubov,  Kubo and  Mori \cite{K57,M65,B61},   has a clear physical meaning since it describes the isothermal susceptibility of the system.
Along the years the Bogoliubov-Kubo-Mori (BKM) metric (under different names: Kubo-Mori scalar product, Bogoliubov inner product, Duhamel two-point function or canonical correlation) has been   intensively studied  in the community of both  physicists and mathematicians , see e.g. \cite {DLS,S93,IJKK82,PT,P94,H97,S00,R09,BT11,B14} and references therein.

  Notably, the Bures metric  $d^2_{B}$  depending on  the fidelity between two density matrices  has
  obtained increasing popularity in the information approach to
  physics, see e.g. \cite{ZVG07,SMK09,AASC10,BT12,T14} and references therein. 
 The  Bures metric $d^2_{B}$ appears under different names: 
 quantum Fisher information (apart from a numerical factor), SLD metric, fidelity susceptibility, etc \cite{P2009,DS14,ZVG07,BC94}.

 Much as for  quantification of the  asymptotic behavior  of the error in the quantum state discrimination problem, one may define the quantum Chernoff metric $d^2_{QC}$, which is expressed in terms of the non-logarithmic variety of the quantum Chernoff bound
 \cite{CMM08,ANSV08,ZMLW14}. The quantum Chernoff metric  also enjoys  increasing popularity in the information approach in physics \cite{AJZ08,IP10,GBMM10}.
It appears  under several different names, among which are (up to a factor of one forth)
the  Hellinger metric \cite{DS14} or the Wigner-Yanase metric \cite{CMM08,ANSV08,ZMLW14}.

Another Riemannian metric in the physical setting, $d^2_{MC}$ (the origin of the subscript MC will become clear in the subsequent consideration), is related to the quadratic fluctuations of a quantum observable 
(the variance of this
observable).
The metric $d^2_{MC}$ was obtained \cite{J86}  as an approximated form of the dispersion of the Umegaki's relative entropy  or
as the Hessian  of the quasi-entropy considered in \cite{P94}.

Here we stress that
all the mentioned  above metrics $d^2_{BKM}$, $d^2_{B}$, $d^2_{QC}$ and $d^2_{MC}$ belong to the large class of the monotone (or contractive) metrics.
In the geometrical approach to statistics proposed by Morozova and $\breve{C}$encov\cite{MC90} the monotone metrics  can be introduced and studied from a unified point of view due to the work of  Petz \cite{P96}.

  In view of Morozova, $\breve{C}$encov and Petz studies
  it is established that there exists  one to one correspondence between  monotone metrics $d^2_f$ and L$\ddot{o}$vner operator monotone functions "$f$" (on the subject of the operator monotone functions see, e.g. the monograph \cite{D74}).
  Each one  of the metrics may be written as a sum of two  contributions; the  classical Fisher-Rao term and a quantum  term that depends on the definition of the inner product that induces the monotone metric on the space of quantum states.  The last is far  to provide  uniqueness. The  consequences  of this fact are subject of considerable interest, see, e.g. \cite{BZ06,H06,PG11,DS14} along with a number of references  therein.
  The relation between operator monotone functions and monotone metrics allows to explorer in deep the relations between the different metrics in the context of quantum statistical mechanics and condensed matter physics.

  The aim of the present study is to show that the isothermal susceptibility (respectively $d^2_{BKM}$)  and the variance of the operator involved through the control parameter (respectively $d^2_{MC}$) may serve as reference metrics  for  the whole class of monotone metrics. In other words, each one of the monotone metrics can be divided into two parts: the $d^2_{BKM}$ (or $d^2_{MC}$) and a constituent in the form of an infinite series of thermodynamic mean values of  products of iterated commutators which are related with the noncommutativity of the problem. It is shown that this  suggests a neat and deep interplay with the linear response theory through the relation with the moments of the Dynamical Structure factor (DSF). If  the Hamiltonian is a linear form  of the generators of a Lie algebra the final result  may be presented in a closed form.

The paper is organized as follows: In Sec.II, using the fact that all monotone Riemannian metrics are characterized by means of operator monotone functions we  derive a generic spectral representation of the monotone Riemannian metrics (see Eq.(\ref{MSP})). We show that this formula provides a sufficiently simple way to derive
different well known metrics, e.g. Bogoliubov-Kubo-Mory, Bures, Chernoff and Morozova-$\breve{C}$encov metrics.  Using this spectral representation, in Sec. III  we obtain new series expansions for
 the monotone Riemmenian  metrics [see Eqs. (\ref{1MSPn}) and (\ref{2MSPn})].  In Sec. IV, the proposed series expansions are checked against  explicit expressions in  a concrete   model. A summary and discussion are given in Sec. V. The Appendix is a compendium of known and not so well known facts about the Bogoliubov-Duhamel inner product, the functionals introduced in \cite{BT11} and their relations to the moments of the dynamical structure factor.

\section{Monotone Riemannian metrics: generic formula }

Following the work of Petz \cite{P96} (see also  \cite{PG11,DS14}) let us recall that a Riemannian metric on  the manifold $\mathcal{D}$ of the density matrices $\rho$ can be written in the form
\begin{equation}
g_{\rho}(X,Y)=\langle X,m_{f}(L_{\rho},R_{\rho})^{-1}(Y)\rangle_{HS},
\end{equation}
where $X,Y$ belong to the  tangent space $T_{\rho}\mathcal{D}$ at $\rho$ of the manifold $\mathcal{D}$, $L_{\rho}(A):=\rho A$ and $R_{\rho}(A):=A\rho$
stand for the left and right multiplication by $\rho$ for $A$ belonging to $T_{\rho}\mathcal{D}$,
\begin{equation}
m_{f}(A,B):=A^{1/2}f(A^{-1/2}BA^{-1/2})A^{1/2}
\end{equation}
is the Kubo - Ando operator means and $\langle AB \rangle_{HS}:=Tr(A^*B)$ is the Hilbert-Schmidt inner product.
Hereafter the subscript $f$ stands for a monotone metric which depends on  the  operator monotone  function  $f(x): f(x^{-1})=f(x)/x$ with $f(1)=1$.
A  Riemannian metric $g_{\rho}$  defines a square distance $d_f^2$ between two infinitesimally close states $\rho$ and $\rho +d\rho$  which in the basis independent form is given by
\begin{equation}
d_f^2=g_{\rho}(d\rho,d\rho).
\label{sdk}
\end{equation}
Note that since the operators $L_{\rho}$ and $R_{\rho}$ commute,    the following relation holds $m_{f}(L_{\rho},R_{\rho})^{-1}=c_{f}(L_{\rho},R_{\rho})$,
where  the symmetric function $c_f(x,y)=c_f(y,x)$,
is called the Morozova-$\breve{C}$encov function.
Recall that  the Morozova-$\breve{C}$encov function has the following presentation
\begin{equation}
c_f(x,y)=
\frac{1}{x f(y/x)}.
\label{CMf}
\end{equation}
The function $c_f(\rho_m,\rho_n)$ obeys the equation $ c_f(tx,ty)=t^{-1}c_f(x,y)$
for any $t \in R$.

 From now for concreteness  we shall consider one-parameter family of density matrices
\begin{equation}
\rho(h) = [Z_N(h)]^{-1}\exp[-\beta {\mathcal H}(h)], \label{roh}
\end{equation}
defined on the family of $N$-particles Hamiltonians of the form
\begin{equation}
{\mathcal H}(h)= T - h S,
\label{ham}
\end{equation}
where the Hermitian operators $T$ and $S$ do not commute in the general case. Here, $h$ is a real (control) parameter, $Z_N(h)= {\mathrm Tr}\exp[-\beta {\mathcal H}(h)]$ is the corresponding partition function and $\beta=(K_B T)^{-1}$ is the inverse temperature.  We assume that the Hermitian operator $T$ has a complete
orthonormal set of eigenvectors $|m\rangle$
with a non-degenerate spectrum $\{T_m\}$; $T|m\rangle = T_m|m\rangle$, where $m=1,2,\dots $ In this basis the zero-field density matrix $\rho :=\rho(0)$ is diagonal:
\begin{eqnarray}
\langle m|\rho (0)|n\rangle &=&\rho_m \delta_{m,n},\qquad
\rho_m:= e^{-\beta T_m}/Z_N(0),
 \qquad m,n = 1,2,\dots .
 \label{zf}
\end{eqnarray}

  Now, let us consider two nearby states $\rho_1=\rho$ and $\rho_2=\rho + d\rho$ produced by infinitesimal changing of the control parameter $h$. We choose $\rho = \rho(0)$
and consider the matrix elements  of  $d\rho$, e.g.  $\langle m|d\rho|n\rangle $, in the basis where  $\rho$ is diagonal.
Thus, for  $\rho=\rho(0)$ with matrix elements (\ref{zf}) through (\ref{sdk})  any monotone Riemannian metric on the set of quantum states (up to a proportionality constant) is presented explicitly as (see, e.g.\cite{BZ06,DS14,ZVG07,Z14}):
   \begin{equation}
  d^2_{f}=\frac{1}{4}\sum_{m}\frac{d \rho_m^2}{\rho_m} + \frac{1}{4}\sum_{m,n,m\neq n}c_f(\rho_m,\rho_n)|\langle m|d\rho|n\rangle|^2.
  \label{CMP}
  \end{equation}

It is however possible to give a more convenient (for calculations) presentation of $d^2_{f}$   by adopting the statistical mechanics viewpoint which will be presented below.

Following \cite{ZVG07} (see also \cite{AASC10,BT12}) for the first term in Eq.(\ref{CMP}) we have
\begin{equation}
\frac{1}{4}\sum_{m}\frac{d \rho_m^2}{\rho_m}=\frac{\beta^2}{4}\langle (\delta S^d)^2\rangle_T,
\label{vrz}
\end{equation}
where $S^d$  is the diagonal part of the operator $S$
and
\begin{equation}
\langle \cdots \rangle_T:= [Z({T})]^{-1}\mathrm{Tr}\{\exp[-\beta {T}]\cdots\}
\end{equation}
denotes  the thermodynamic mean value.
A next step consists in using the following  relations (see, e.g.\cite{ZVG07,V10}):
 \begin{equation}
 \langle m|d\rho|n\rangle= \langle m|dn\rangle (\rho_n-\rho_m),\qquad m\neq n,
 \label{ur1}
 \end{equation}
 and
 \begin{equation}
  \langle m|dn\rangle=\frac{\langle m|\partial {\mathcal H}(h)|n\rangle}{T_n - T_m},\qquad
  \partial {\mathcal H}(h)\equiv\frac{\partial {\mathcal H}(h)}{\partial h}= - S.
  \label{ur2}
  \end{equation}
  By means of Eqs.(\ref{vrz}), (\ref{ur1}) and (\ref{ur2}),
   Eq.(\ref{CMP}) takes the more convenient form
  \begin{eqnarray}
d^2_{f}=\frac{\beta^2}{4}\left\{\langle (\delta S^d)^2\rangle_T
 + \sum_{m,n,m\neq n}c_f(\rho_m,\rho_n)\right.
\left. \left(\frac{\rho_n -\rho_m}{\ln \rho_n -\ln \rho_m}\right)^2|\langle m| S|n\rangle |^2\right\}.
\label{MSP}
\end{eqnarray}
This expression is a basic element in our further calculations.

There are so many monotone metrics $d^2_f$ as $f$'s are.
Ones that are  mostly proposed in the literature operator monotone functions are as follows \cite{P96}:

  \begin{eqnarray}
f_{Har}(x)=\frac{2x}{x+1}, f_{BKM}(x)=\frac{x-1}{\ln x},
f_{WY}(x)=\frac{(1+\sqrt{x})^2}{4},
f_B(x) = \frac{x+1}{2},
f_{MC}(x)=\left(\frac{x-1}{\ln x}\right)^2\frac{2}{1+x}.
\label{uga}
\end{eqnarray}
The function $f_{Har}(x)$ is the minimal, while $f_{B}(x)$ is the maximal operator monotone functions on $[0,+\infty)$. The former  defines a metric  known as RLD metric while the last one gives rise to the SLD   metric (named also Bures metric  or fidelity susceptibility). The   function $f_{BKM}(x)$  leads to the  Bogoliubov-Kubo-Mori  metric.
 The function $f_{WY}(x)$ is associated to the Wigner-Yanase metric (or an equal ground the name quantum Chernoff  metric is reasonable ).  The function $f_{MC}(x)$
 was first conjectured in the paper \cite{MC90}, which explains the subscript $MC$.  Its matrix monotonicity was proved in \cite{P96}.

It is instructive to consider the generic formula Eq. (\ref{MSP}) in the concrete cases of the operator monotone functions (\ref{uga}).

\subsection{Bogoliubov - Kubo - Mori metric}

The derivation  of the BKM metric and its physical meaning as susceptibility ware discussed earlier in \cite {DLS,S93,PT,P94,H97,S00,R09,BT11,B14}. In order to introduce the needed definitions  we shall give our slightly different derivation based on the notion of the Bogoliubov-Duhamel (or Kubo-Mori) inner product (see the Appendix).
The corresponding Morozova-$\breve{C}$encov  function (everywhere below we shall use the subscript of the function $f$ instead of the very function) is
\begin{equation}
c_{BKM}(\rho_m,\rho_n)=\left(\frac{\ln \rho_n - \ln \rho_m}{\rho_n-\rho_m}\right).
\label{mnrB}
\end{equation}

From the other hand, let us consider the spectral representation of the  Bogoliubov-Duhamel   inner product \cite{B61,DLS,PT,BT11} (see also the Appendix):

\begin{eqnarray}
F_{0}(S;S):=\frac{1}{2}\sum_{m,n, m\not=n} |\langle n|S|m \rangle|^2\frac{\rho_n
-\rho_m}{X_{mn}}+ \sum_{n}\rho_n
\langle n|S|n\rangle|^2,
\label{M5111}
\end{eqnarray}
where for  convenience   the quantity
\begin{equation}
X_{mn}=\frac{1}{2}(\ln \rho_n-\ln \rho_m)\equiv\frac{\beta}{2}(T_m-T_n)
\label{newq}
\end{equation}
 is introduced. It can be shown by simple algebra that
\begin{equation}
F_{0}(\delta S;\delta S)=F_{0}(S;S)-|\langle S \rangle_T|^2, 
\end{equation}
where
 $\delta S = S - \langle S\rangle_T$.
Since by definition
\begin{equation}
\langle (\delta S^d)^2 \rangle_T :=\sum_m \rho_{m}\langle m|S|m \rangle^2 - \langle S\rangle_T^2
\label{sdr},
\end{equation}
where $\delta S^d = S^d - \langle S^d\rangle_T$ ,
we can present Eq.(\ref{M5111}) in the form
\begin{equation}
 \frac{1}{2}\sum_{m,n,m\neq n}|\langle m|S|n\rangle |^2\frac{\rho_m -\rho_n}{X_{nm}}= F_{0}(\delta S;\delta S)-\langle (\delta S^d)^2\rangle_T .
\label{BD1}
\end{equation}

In view of Eqs. (\ref{mnrB}) and (\ref{BD1})  from  Eq.(\ref{MSP}),
we obtain
\begin{eqnarray}
d^2_{BKM}=\frac{\beta^2}{4}F_{0}(\delta S;\delta S).
\label{dBKM}
\end{eqnarray}
Recall that alternatively we have the well known relation (see, e.g. \cite{W67})
\begin{equation}
F_{0}(\delta S;\delta S)=\int_0^1d\tau \left\langle \left[\delta S(\tau)\delta S\right]\right\rangle_T=\frac{1}{\beta^{2}}\frac{\partial^2\ln Z(h)}{\partial^2 h}\vert_{h=0}.
\label{vaz}
\end{equation}
Using the definition of the isothermal susceptibility with respect to the field $h$:
\begin{equation}
\chi_{h=0}^{T}:=\frac{1}{\beta}\frac{\partial^2 \ln Z(h)}{\partial h^2}|_{h=0},
\end{equation}
   the following result emerges
\begin{equation}
d^2_{BKM}=\frac{\beta}{4}\chi_{h=0}^{T}.
\label{er}
\end{equation}
We shall follow the above  recipe  in order to study the relations between different metrics.

\subsection{Morozova-$\breve{C}$encov  metric}
We shall show   that the operator monotone function
$f_{MC}(x)=\left(\frac{x-1}{\ln x}\right)^2\frac{2}{1+x}$ makes sense in physical applications.
The corresponding Morozova-$\breve{C}$encov  function
\begin{equation}
c_{MC}(\rho_m,\rho_n)=\left(\frac{\ln \rho_n - \ln \rho_m}{\rho_n-\rho_m}\right)^2\frac{\rho_n+\rho_m}{2}
\label{mnr}
\end{equation}
yields the  metric $d^2_{MC}$ already discussed in the Introduction. Since to our knowledge  appropriate  citation of this fact is unknown a simply deviation is presented.
By inserting (\ref{mnr}) into (\ref{MSP}),
 we obtain
\begin{equation}
d^2_{MC}=\frac{1}{4}\sum_{m}\frac{d \rho_m^2}{\rho_m} + \frac{\beta^2}{8}\sum_{m,n,m\neq n}|\langle m|S|n\rangle |^2(\rho_n +\rho_m).
\label{MSP1}
\end{equation}
Now, using  the  relation
\begin{equation}
\langle S^2 \rangle_T=\frac{1}{2}\sum_{m,n}|\langle m|S|n\rangle|^2(\rho_{m}+\rho_n)
\end{equation}
and Eqs. (\ref{vrz}) and
(\ref{sdr})  we get the final result
\begin{equation}
d^2_{MC}=\frac{\beta^2}{4}{\langle(S-\langle S \rangle_T)^2\rangle_T}.
\label{dMC}
\end{equation}
Note that in the case of commuting operators $T$ and $S$, according to Eq.(\ref{vaz}) both metrics
$d_{BKM}^2$ and $d_{MC}^2$ coincidece, as it must be.

\subsection{Bures metric}
There are  different ways to obtain the analytical expression of the Bures metric $d_{B}^2$.
It  may be obtained from the infinitesimally close form of the Bures distance.
For any two states $\rho_1$ and $\rho_2$  the Bures distance  can be expressed in terms of the Uhlmann-Jozsa fidelity:
\begin{equation}
{\cal F}(\rho_1,\rho_2) = \mathrm{Tr}\sqrt{\rho_1^{1/2} \rho_2 \rho_1^{1/2}}, \label{defFidel}
\end{equation}
as
\begin{equation}
D_{B}(\rho_1,\rho_2)=\sqrt{2-2{\cal F}(\rho_1,\rho_2)}
\label{Bures}.
\end{equation}
Thus   $d_{B}^2$ is obtained as the leading term in the expansion of Eq.(\ref{Bures}).
 Among the operator monotone functions introduced by Morozova $\check{C}$encov and Petz it is associated with the maximal operator monotone function $f_B(x)=\frac{x+1}{2}$. By setting it in Eq.(\ref{MSP}) we obtain the result
\begin{equation}
d^2_{B}=\frac{1}{4}\sum_{m}\frac{d \rho_m^2}{\rho_m} +\frac{\beta^2}{2}\sum_{m,n, m\not=n} \frac{|\langle n|S|m \rangle|^2}{|\ln \rho_m -\ln \rho_n|^2}
 \frac{(\rho_n-\rho_m)^2}{\rho_n+\rho_m},
\label{FSus0}
\end{equation}
(see, e.g. \cite{ZVG07,AASC10,BT12,T14} and comments therein).

\subsection{Chernoff/Wigner-Yanase metric}
The quantum Chernoff metric
is the infinitesimally close form
 of the Chernoff distance:
 \begin{equation}
   D_{QC}(\rho_1,\rho_2)=1-Q(\rho_1,\rho_2),
   \label{Chern}
   \end{equation}
where
 \begin{equation}
   Q(\rho_1,\rho_2)=\min_{0\leq s \leq 1}Tr(\rho_1^{1-s}\rho_2^s)
   \label{nlChern}
   \end{equation}
is known as the nonlogarithmic variety of the quantum Chernoff bound \cite{ACM07}. It can be written as \cite{ZVG07,ACM07,CMM08,ZMLW14,AJZ08}
\begin{eqnarray}
 d^2_{QC}=\frac{1}{8}\sum_{m}\frac{d \rho_m^2}{\rho_m} +
 \frac{\beta^2}{2}\sum_{m,n, m\not=n}
\frac{|\langle n|S|m \rangle|^2}{|\ln\rho_m -\ln\rho_n|^2}
(\rho_m^{1/2} - \rho_n^{1/2})^2.
\label{FiSus3}
 \end{eqnarray}
The metric (\ref{FiSus3}) coincides up to a factor $1/2$ with  the Wigner-Yanase metric \cite{GI03}, i.e. $d^2_{QC}=(1/2) d^2_{WY}$, which
may be obtained  from Eq.(\ref{MSP}) with the help of the operator monotone function $f_{WY}(x)=\frac{(1+\sqrt{x})^2}{4}$.

 Here, the following comment is in order. In terms of Green's functions a finite-temperature generalization of the fidelity susceptibility was proposed in \cite{AASC10}.  This quantity is quite  different from the Bures-Uhlmann  fidelity susceptibility (i.e. $d^2_{B}$) and we denoted it as $\chi^T_{F}$. The only similarity between both metrics is that they have  the same $T=0$ limit. The fidelity susceptibility $\chi^T_{F}$  is obtained as the first nonvanishing term in the expansion of the "fidelity"
\begin{equation}
{\cal F}_{T}(\rho_1,\rho_2)=\sqrt{Tr(\rho_1^{1/2}\rho_2^{1/2})}.
\end{equation}
For for more details see ref. \cite{S10}. It is worth noting, that the spectral representation of  $\chi^T_{F}$ is presented  by Eq. (14) in ref.\cite{AASC10}.
It is
simply to check  that, in our notations, it emerges
exactly as the quantum Chernoff metric $d^2_{QC}$, Eq.(\ref{FiSus3}). Since, the computational problems in studying $\chi^T_{F}$ can be efficiently tackled by the quantum Monte Carlo approach \cite{AASC10} this fact transfers  new computational possibilities  in the field concerning the quantum Chernoff metric.

\section{BKM and MC metrics as reference metrics }
In order to quantify   the deviation of the one of monotone Riemannian metric $d^2_f$ from the $d^2_{BKM}$ or $d^2_{MC}$  we shall recast Eq. (\ref{MSP}) in a form suitable for further elaborations.
\subsection{BKM metric}
 In order to discuss the deviation of a metric from the BKM metric, we need to rewrite  Eq. (\ref{MSP}) in the form:
\begin{eqnarray}
d^2_{f}=\frac{1}{4}\sum_{m}\frac{d \rho_m^2}{\rho_m} + \frac{\beta^2}{4}\sum_{m,n,m\neq n}g_f\left(X_{mn}\right)
\frac{\rho_n -\rho_m}{\ln \rho_n -\ln \rho_m}|\langle m| S|n\rangle |^2,
\label{MSPn}
\end{eqnarray}
where we introduced the family of functions
\begin{equation}
g_f(x):=\frac{e^{2x}-1}{2xf(e^{2x})}.
\label{gf}
\end{equation}
These functions satisfy (due to the symmetry $xf(x^{-1})=f(x)$) the condition
$g_f(x)=g_f(-x)$ and $g_f(0)= 1$.
In the considered particular cases    of $f$'s one is led to

\begin{eqnarray}
g_{Har}\left(x\right)=\frac{\sinh 2x}{2x}, g_{BKM}\left(x\right)=1,
g_{WY}\left(x\right)=\frac{\tanh\frac{1}{2}x}{\frac{1}{2}x},
 g_B\left(x\right)=\frac{\tanh x}{x},
 g_{MC}\left(x\right)=\frac{x}{\tanh x}.
\label{uga1}
\end{eqnarray}
(We use argument   $X_{mn}$ in the function $g_f(.)$  instead of $\frac{\rho_m}{\rho_n}$  in the  rhs of (\ref{MSPn}),
 remembering the relation Eq.(\ref{newq}).)
 The presentation (\ref{MSPn}) is convenient for two reasons.

 First, using the well known inequalities between hyperbolic functions one obtains
\begin{eqnarray}
g_{Har}\left(X_{mn}\right)\geq g_{MC}\left(X_{mn}\right) \geq g_{BKM}\left(X_{mn}\right)(=1)
\geq g_{WY}\left(X_{mn}\right)
\geq g_B\left(X_{mn}\right).
\label{CMT1}
\end{eqnarray}
From the above inequalities immediately follow corresponding inequalities between different
metrics.

 Second, since functions $g_{f}\left(X_{mn}\right)$ are even functions they have  series expression with only even degrees of $X_{mn}$.
Thus,   functions $g_{f}\left(X_{mn}\right)$ can be expressed  in a unified fashion  as a formal series
\begin{equation}
g_{f}\left(X_{mn}\right)=1 + \sum_{l=1}^{\infty}a_{2l-1}(f) (X_{mn})^{2l}.
\label{use}
\end{equation}
We hope no confusion will arise by using the subscript $2l-1$ instead of $2l$ in order to enumerate the coefficients $a$'s in the above  expansion.
Particulary, in the case of Bures metric from the series expansion of the function $x^{-1}\tanh x$ it is easy to obtain
\begin{equation}
a_{2l-1}(B)=\frac{2^{2l+2}(2^{2l+2}-1)}{(2l+2)!}B_{2l+2},
\end{equation}
where $B_{2l+2}$ are the Bernoulli numbers.
The separation in coefficients $a_l(f)$  the very dependence on functions $f$   allows to introduce the terms
\begin{equation}
2^{-2l}a_{2l-1}(f)(\ln \rho_n-\ln \rho_m)^{2l-1}(\rho_n -\rho_m)|\langle m| S|n\rangle |^2
\label{mn}
\end{equation}
in the expansion Eq.(\ref{MSPn}). Remarkably, we shall show that these terms generate well known thermodynamic mean values.
The presentation (\ref{MSPn}) provides  the role of Bogoliubo-Kubo-Mory  metric as a reference metric.  Recall that
 Eq. (\ref{MSPn})  simply presents the definition of $d^2_{BKM}$ if we choose $g_f\left(X_{mn}\right) =g_{BKM}\left(X_{mn}\right)=1$.
Thus $g_f\left(X_{mn}\right) \neq 1$  measures the deviation of the corresponding metric $d^2_f$ from the Bogoliubo-Kubo-Mory metric.

\subsection{MC metric}
One may use an alternative presentation of $d_{f}^2$ instead of Eq. (\ref{MSPn}):
\begin{eqnarray}
d^2_{f}=\frac{1}{4}\sum_{m}\frac{d \rho_m^2}{\rho_m} + \frac{\beta^2}{4}\sum_{m,n,m\neq n}\hat{g}_f\left(X_{mn}\right)
\frac{\rho_n +\rho_m}{2}|\langle m| S|n\rangle |^2,
\label{MSPM}
\end{eqnarray}
where
\begin{equation}
\hat{g}_f\left(x\right):=g_f\left(x\right)\frac{\tanh x}{x}.
\label{hat}
\end{equation}
Now,
as a reference metric one can use $d^2_{MC}$. If we set $f=f_{MC}$ in Eq.(\ref{hat}) then $\hat{g}_{MC}\left(X_{mn}\right)=1$
and Eq.(\ref{MSPM}) reduces to the definition of $d^2_{MC}$. Hence, functions $ \hat{g}_{f}$ serve as a measure of deviation of the corresponding  $d^2_f$'s from $d^2_{MC}$.

For the  functions $f$ presented in  (\ref{uga}), the application of  definitions (\ref{uga1}) in (\ref{hat}) yields
\begin{eqnarray}
 &&\hat{g}_{Har}(x)=\frac{2(\cosh 2x-1)}{(2x)^2},\quad  \hat{g}_{BKM}\left(x\right)=\frac{\tanh x}{ x}, \nonumber\\
 && \hat{g}_{MC}\left({x}\right)=1,\quad
 \hat{g}_{WY}(x)=\frac{2[1-(\cosh x)^{-1}]}{x^2},\nonumber\\
  && \hat{g}_B(x)=\left[\frac{\tanh x}{x}\right]^2.
 \label{uga2}
\end{eqnarray}
The even functions Eq.(\ref{hat}) can be presented  as a formal series
\begin{equation}
\hat{g}_{f}\left(X_{mn}\right)=1 + \sum_{l=1}^{\infty}a_{2l}(f) (X_{mn})^{2l},
\label{use1}
\end{equation}
which introduces in the summand of the expansion Eq.(\ref{MSPM}) the terms
\begin{equation}
 2^{-(2l+1)}a_{2l}(f)(\ln \rho_n-\ln \rho_m)^{2l}(\rho_n +\rho_m)|\langle m| S|n\rangle |^2.
\label{ms}
\end{equation}
These terms  also can be expressed as some thermodynamical mean values (c.f. with Eq.(\ref{mn})).
In the next subsection we shall show   that $d^2_{MC}$   appears as a first term of the obtained series representation of the monotone Riemannian  metrics.

\subsection{Series presentations in terms of iterated commutators}

Let us consider the functionals
\begin{eqnarray}
F_{n}(S;S) :=  \sum_{ml}|\langle m|S|l \rangle |^{2}\frac{|\rho_{l}
- (-1)^n \rho_{m}|}{|\ln \rho_l - \ln \rho_m|^{n-1}},\quad
 n=0,1,2,....
\label{BT10abc}
\end{eqnarray}
These functionals have been introduced in \cite{BT11} (see also the Appendix) as an useful tool to obtain different thermodynamic inequalities.
 By using the definition (\ref{BT10abc}), one can present the terms (\ref{mn}) and (\ref{ms}) in the form
\begin{equation}
2^{-2l}a_{2l-1}(f)F_{2l}(S;S)
\label{ev}
\end{equation}
and
\begin{equation}
2^{-(2l+1)}a_{2l}(f)F_{2l+1}(S;S),
\label{od}
\end{equation}
respectively.  Now, the Eq.(\ref{MSPn})  reads
\begin{equation}
d^2_{f}=d^2_{BKM} + \frac{\beta^2}{4}\sum_{l=1}^{\infty}2^{-2l}a_{2l-1}(f)F_{2l}(S;S).
\label{1MSPn}
\end{equation}
Similarly,  the Eq.(\ref{MSPM}) reads
\begin{equation}
d^2_{f}=d^2_{MC} + \frac{\beta^2}{4}\sum_{l=1}^{\infty}2^{-(2l+1)}a_{2l}(f)F_{2l+1}(S;S).
\label{2MSPn}
\end{equation}
Indeed, the first terms in Eqs.(\ref{1MSPn}) and (\ref{2MSPn}) are immediate consequences of the definitions (\ref{dBKM}) and (\ref{dMC}), while the summands follow from the terms
(\ref{ev}) and (\ref{od}), respectively. The key advantage of both formulas is  that
in the basis independent form functionals $F_{n}(S;S)$ have  the following  presentations (see the Appendix)
\begin{equation}
F_{n}(S;S)=2(-1)^{n+1}\beta^{n-1}\langle R_{n-1}R_{0}\rangle_T,
\label{Nf}
\end{equation}
where the notion of iterated  commutators
\begin{eqnarray}
&&R_{0}\equiv R_{0}(S)\equiv S,\quad  R_{1}\equiv R_{1}(S) :=[T,S],\;
\dots, \; \nonumber\\
&& R_{n} \equiv R_{n}(S):= [T,R_{n-1}(S)],\quad n=0,1,2,...,
\label{R1a}
\end{eqnarray}
is introduced. (Note that  nested commutators is also frequently used term.)  It is worthwhile to emphasize the relation of Eq. (\ref{Nf}) with the moments of the dynamical structure
factor $M_{p}(S)$  (see the Appendix)
\begin{equation}
F_{n}(S;S)= 2\beta^{n-1}M_{n-1}(S),\quad n=0,1,2,...
\label{VRNTa}
\end{equation}
With the help of Eq.(\ref{VRNTa}) one obtains from Eqs. (\ref{1MSPn}) and (\ref{2MSPn}) the result
\begin{equation}
d^2_{f}=  d^2_{BKM} +  \frac{\beta^2}{4}\sum_{l=1}^{\infty}\left(\frac{\beta}{2}\right)^{2l-1} a_{2l-1}(f) M_{2l-1}(S). 
\label{1MSPna}
\end{equation}
and
\begin{equation}
d^2_{f}=d_{MC}^2 + \frac{\beta^2}{4}\sum_{l=1}^{\infty}\left(\frac{\beta}{2}\right)^{2l} a_{2l}(f) M_{2l}(S). 
\label{1MSPnb}
\end{equation}
This result attributes thermodynamical meaning  to pure information theory ingredients.

The series representations (\ref{1MSPn}) and (\ref{2MSPn})  yield a proper definition of a monotone Riemannian metric provided the corresponding convergence condition is fulfilled.
In order to remove the restrictions imposed by the convergence conditions one needs to perform an analytic extension of the obtained formulas.
This issue in the particular case of the fidelity susceptibility (Bures metric) has been examined in the ref.\cite{T14} by the examples of several popular models.

The appearance of the iterated commutators (terms with $n>1$) is a reminiscence of the well known Feynman's disentangling procedure \cite{W67,V68}.

In the next Section we shall  demonstrate that this allows  the underlaying symmetry of the Hamiltonian  to be  efficiently explored and a closed form  of the functionals (\ref{BT10abc})
to be obtained.

\section{Application to a model}

Let us consider the Hamiltonian  \cite{LYZ10,Z13}:
\begin{equation}
{\mathcal H}(h)=k\omega\left(Q^0_k-\frac{1}{k^2}\right)+h\sqrt{k^k}(Q^+_k + Q^-_k),\quad k=1,2,...,
\label{HAQ}
\end{equation}
where  $Q^\pm_k$ are operators obeying
 the commutation relations
\begin{equation}
[Q^0_k,Q^{\pm}_k]= \pm Q^{\pm}_k,\qquad [Q^+_k,Q^{-}_k]= \Phi_k(Q^0_k)-\Phi_k(Q^0_k-1),
\label{KR1}
\end{equation}
with the structure function
\begin{equation}
\Phi_k(Q^0_k)=-\Pi_{i=1}^{k}\left(Q^0_k+\frac{i}{k}-\frac{1}{k^2}\right)
\end{equation}
being a $k^{th}$-order polynomial in $k$.
The Hamiltonian (\ref{HAQ})
is employed in various physical problems (for definitions  and a partial list of references, see \cite{Z13,LYZ10}).

In this case the proposed approach  is very effective since the iterative commutation
 between  $T=k\omega\left(Q^0_k-\frac{1}{k^2}\right)$ and $S=\sqrt{k^k}(Q^+_k + Q^-_k)$
implies  some periodic operator structures  after a finite number of steps \cite{T14}
 \begin{equation}
R_n=(-1)^nR^+_{n}=\alpha^n[Q^+_k + (-1)^nQ^{-}_k], \quad (Q^-_k)^+=Q^{+}_k ,
\label{sf1}
\end{equation}
 indicating an analytical expression  as a function of $n$.
The parameters $k$ and $\omega $  enter in the c-number $\alpha=(k\omega)^k \sqrt{k^k}$ . Thus, the  obtained series expansions Eqs. (\ref{1MSPn}) and (\ref{2MSPn}) can be used in a rather simple  way to obtain  closed-form expressions.

The polynomial
algebra of degree $k-1$  defined  by Eqs.(\ref{KR1})
has  the following one-mode boson realization \cite{LYZ10}:
\begin{equation}
 Q^+_k= \frac{1}{(\sqrt{k})^k} (b^+)^k,\quad  Q^-_k = \frac{1}{(\sqrt{k})^k} b^k.
\label{pZ}
\end{equation}
In terms of Eqs. (\ref{pZ}) the Hamiltonian of the model takes the more familiar form \cite{Z13}
\begin{equation}
{\mathcal H}(h)=\omega b^+b + h [(b^+)^k + b^k],\qquad \omega >0,\quad k=1,2,3,...
\label{HAk}
\end{equation}
where bosonic operators $b, \;b^+$   obey  the canonical commutation
relations.

It is worse noting that the k = 1 and
k = 2 cases of (\ref{HAk}) give the Hamiltonians of the displaced and single-mode squeezed
harmonic oscillators, respectively. The Hamiltonian (\ref{HAk}) for $k=2$ is also known as Lipkin-Meshkov-Glick (LMG) model in the Holstein-Primakoff single boson representation (see e.g. \cite{Gu10} and refs. therein)
and all the result obtained here can be related to this field.

Inserting the  expressions  $R_0$ and $R_{2n-1}$ in (\ref{Nf}), we obtain
\begin{eqnarray}
F_{2n}(S;S)=-2(k\beta\omega)^{2n-1}{\mathcal K}(k),\quad n=0,1,2,...,\quad k=1,2,...,
\label{K}
\end{eqnarray}
where
\begin{eqnarray}
{\mathcal K}(k)=k^k\langle[Q^{+}_k - Q^-_k][Q^{+}_k + Q^-_k]\rangle_T,\quad
k=1,2,...
\label{MKa}
\end{eqnarray}
Inserting expressions of $R_0$ and $R_{2n}$ in (\ref{Nf}), we obtain
\begin{eqnarray}
&&F_{2n+1}(S;S)=2(k\beta\omega)^{2n}{\mathcal L}(k),\nonumber\\
&& n=0,1,2,...,\quad k=1,2,...,
\label{Kdr}
\end{eqnarray}
where
\begin{equation}
{\mathcal L}(k)=k^k\langle[Q^{+}_k + Q^-_k]^2\rangle_T,\qquad k=1,2,...
\label{MKb}
\end{equation}
Evaluation of the correlation functions Eqs. (\ref{MKa}) and (\ref{MKb}) with the quadratic Hamiltonian T is now straightforward.
The results for $k=1$ and $k=2$ are:
\begin{eqnarray}
{\mathcal K}(1) &=&-1,\quad {\mathcal L}(1)=2n+1, \nonumber\\
{\mathcal K}(2) &=& -2(2n+1),\quad {\mathcal L}(2)=4n^2,
\end{eqnarray}
where $n=(e^{\beta \omega} - 1)^{-1}$.
With the help of Eqs.(\ref{K}) and (\ref{Kdr}), Eqs.(\ref{1MSPn}) and (\ref{2MSPn})  may be recast in the form
\begin{equation}
d^2_{f}=d^2_{BKM} + \frac{\beta^2}{4}\left(\frac{k\beta\omega}{2}\right)^{-1}\left[1-g_{f}\left(\frac{k\beta\omega}{2}\right)\right]{\mathcal K}(k)
\label{nu1}
\end{equation}
and
\begin{equation}
d^2_{f}=d^2_{MC} - \frac{\beta^2}{4}\left[1- \hat{g}_{f}\left(\frac{k\beta\omega}{2}\right)\right]{\mathcal L}(k),
\label{nu2}
\end{equation}
respectively.
The relation (\ref{nu1}) on the particular example of  $f_{B}$  was  found earlier in \cite{T14}. Here we are able to consider the whole class of  the monotone Riemannian metric.

\section{Summary and Conclusions}

In this paper, we study model systems described by a Hamiltonian comprising the non commuting operators $T$ and $S$ via  monotone
Riemannian metrics.
The formulas Eqs. (\ref{1MSPn}) and (\ref{2MSPn}) are the key results of our study.
They present the  monotone Riemannian metric as a series in terms of the earlier introduced \cite{BT11} functionals $F_n(S;S)$  (see also the Appendix).
These are defined as a rather complicated double sum which however may be written  in a basis independent form as thermodynamic mean values of n-times iterated commutators
 between  $T$ and $S$.  The last provides significant computational advantage if the Hamiltonian is a linear form of the generators of some Lie algebra.
  In this case  $F_{n}(S;S)$ should be  obtained in a closed-form expression as a function of $n$. Recall  that the lowest-ordered functionals $F_0(S;S)$ and $F_{1}(S;S)$ are
  the Bogoliubov-Duhamel inner product and the symmetrized thermodynamic mean value of the operator $S$, respectively.  Then $F_0(\delta S;\delta S)$ is proportional to the  BKM metric given by Eq.(\ref{er}),
  while $F_{1}(\delta S;\delta S)$ is proportional to the MC metric given by Eq.(\ref{dMC}). These quantities  are the starting point in our expansions  Eqs. (\ref{1MSPn}) and (\ref{2MSPn}).

From the linear response theory it is well known that the iterated commutators
 of an observable $S$ with the Hamiltonian  are related to the moments of the dynamic structure factor (DSF)  through some sum rules \cite{F80,PS04}.
This allows to present our results in  an alternative form, Eqs. (\ref{1MSPna}) and (\ref{1MSPnb}).
A look at the above formulas shows that the dependance on the operator monotone function $f$ is only in the coefficients in front of the moments.
 If one aims to characterize a monotone metric on this setting, then moments of all orders
(odd or even) should be considered together.

Formulas Eqs. (\ref{1MSPna}) and (\ref{1MSPnb}) do not provide    computational
benefits per se.
 However they are very useful and informative because they transfer  the information geometry   problems into the realm of condensed mater physics  with its wealth of   methods for computing the DSF and its moments.
For example, an instructive illustration  may present the study  \cite{C15}, where an application of appropriate (physical) approximations to Hamiltonian described  one-component Coulomb plasma in thermodynamic equilibrium leads to  explicit formulas for the arbitrary integer moments of the DSF expressed in terms  of simple functions.

The series representation (\ref{1MSPn}) (or (\ref{2MSPn}))  may be regarded as a proper definition of the considered monotone Riemannian metric provided the corresponding convergence condition is fulfilled which can be checked on the framework of concrete models.

We demonstrated our approach in an example with a Hamiltonian, expressed in terms of  the generators of a polynomial deformation
Lie algebra, Eq.(\ref{HAQ}), employed in various physical problems. It is shown that in this case, the infinite set of  the moments of all orders
(odd and even) can be written in a closed form,
Eqs.(\ref{K}) and Eq.(\ref{Kdr}), for the all monotone Riemannian metrics.
Besides being of interest for its own sake the presented result  may also be considered as a contribution to the linear response theory.

At the end the following comment is in order. The implication of the immanent  relations between the zero-temperature  fidelity susceptibility  and the moments of the DSF  recently
has been demonstrated in ref. \cite{YH15}. However the elucidation of  the role of the moments $M_{p}(S)$ (named as p-order generalized fidelity susceptibility) has been considered in quite different  context.

\section*{Acknowledgments}

This work was supported by EU FP7 INERA project grant agreement number 316309.

\section*{Appendix: The Bogoliobov-Duhamel inner product, moments of the dynamic structure factor and sum rules}

Let us $S$ is an arbitrary operator.  The Bogoliubov- Duhamel inner product is defined us \cite{B61,DLS,S93,PT,BT11}:

\begin{equation}
(S;S)_T=\int_0^1d\lambda K_{SS}(\lambda),\quad
\label{BDIP}
\end{equation}
where
\begin{equation}
K_{SS}(\lambda):=\left\langle S^+(\lambda)S\right\rangle_T, \quad S^+(\lambda):=e^{\lambda\beta T}S^+e^{-\lambda\beta T},
\label{sp1}
\end{equation}
and
$$\langle \cdots \rangle_T:= [Z({T})]^{-1}\mathrm{Tr}\{\exp[-\beta {T}]\cdots\}$$
denotes the thermodynamic mean value. We warm the reader that definitions differ from (\ref{BDIP}) by factor $\beta$ and/or by involving on the first place in the  Eq.(\ref{sp1}) the operator
$S$ instead of its adjoint $S^+$ exist in the literature.

With the help of the Kubo identity (see, e.g.\cite{W67})
\begin{equation}
[S,e^{-\beta T}]=-\beta \int_0^1d\tau e^{-\beta(1-\tau)T}[S,T]e^{-\beta\tau T},
\end{equation}
 the following useful formula can be verified:
\begin{equation}
\beta(S;[T,S])_T=\langle[S^+,S]\rangle_T,
\label{KBD}
\end{equation}
where $S$ and $T$ are arbitrary operators.

The definition of the Bogoliubov- Duhamel inner product [Eq. (\ref{BDIP})] can be presented in terms of the spectral representation in which $T$ is diagonal.
Let us assume that the Hermitian operator $T$ has a complete orthonormal set of eigenstates $|l\rangle$ and eigenvalues $T_l$; $T|l\rangle= T_l|l\rangle$. Then, in the basis of the eigenvectors of the Hamiltonian
$T$ we have the alternative form
\begin{eqnarray}
(S;S)_{T}:=\frac{1}{2}\sum_{m,l, m\not=l} \frac{\rho_l
-\rho_m}{X_{ml}} |\langle m|S|l \rangle|^2 + \sum_{l}\rho_l
|\langle l|S|l \rangle|^2 ,
\label{M511}
\end{eqnarray}
where the notations $\langle l|\rho|l\rangle=\rho_l$ and $X_{ml}= 2^{-1}\beta  (T_m-T_l)$ are introduced.

  It is convenient to consider the spectral representation (for more details and  history remarks, see e.g.\cite{R09})
\begin{equation}
K_{SS}(0)=\int_{-\infty}^{\infty}d\omega Q_{S}(\omega),
\label{SS0}
\end{equation}
where
\begin{equation}
Q_{S}(\omega)=[Z(T)]^{-1}\sum_{m,n}e^{-\beta T_m}|\langle n|S|m \rangle|^2\delta(\omega -\omega_{nm}) \geq 0,
\label{sf}
\end{equation}
is the dynamic structure factor (DSF) relative  to the operator $S$,
where $\omega$ is a real frequency and $\omega_{nm}=T_n-T_m \quad (\hbar=1)$. DSF obeys the relation
\begin{equation}
Q_{S}(\omega)=e^{\beta\omega}Q_{S^+}(-\omega),
\label{dbp}
\end{equation}
known as the principle  of detailed balancing.

 Correspondingly  the spectral representation of  $K_{SS}(\lambda)$ is given by Luttinger \cite{L66}
\begin{equation}
K_{SS}(\lambda)=\int_{-\infty}^{\infty}d\omega Q_{S}(\omega;\lambda),\qquad Q_{S}(\omega;\lambda)=Q_{S}(\omega)e^{\lambda \beta \omega}.
\label{Bsr}
\end{equation}
Hence, as follows from Eqs. (\ref{BDIP}) and (\ref{Bsr}) we get
\begin{equation}
(S;S)_T = \int_{-\infty}^{\infty}d\omega\frac{e^{\beta\omega}-1}{\beta\omega}Q_{S}(\omega),
\label{Bpip}
\end{equation}
which is the  original expression obtained by Bogoliubov\cite{B61}. For the further consideration it is useful to introduce the moments of the DSF
\begin{equation}
M_{p}(S):=\int_{-\infty}^{+\infty}d\omega \omega^{p}Q_{S}(\omega),\quad p=-1,0,1,2,...
\label{dm1}.
\end{equation}
 Then another alternative form of Eq.(\ref{BDIP}) is
\begin{equation}
(S;S)_T=\beta^{-1}[M_{-1}(S) + M_{-1}(S^+)].
\label{m1}
\end{equation}
 Eq.(\ref{m1}) directly follows from Eqs. (\ref{sf}), (\ref{dbp}) and (\ref{dm1}).

In our paper \cite{BT11} the
functionals $F_{n}(S;S), n=0,1,2,...$
\begin{equation}
F_{n}(S;S) := 2^{n-1} \sum_{ml}|\langle m|S|l \rangle |^{2}|\rho_{l}
- (-1)^n \rho_{m}| . |X_{ml}|^{n-1},
\label{BT10ab}
\end{equation}
have been introduced as a generalization of the Bogoliubov- Duhamel inner product [Eq.(\ref{M511})]. Some applications of these functionals   have been demonstrated in  \cite{BT11,BT12,T14}.
Since $F_{0}(S;S)=(S;S)_{T}$ hereafter in the text  we shall use this notation for the Bogoliubov-Duhamel inner product. Recall that $F_1(S;S)=\langle S^+S + SS^+ \rangle_T$.
 Using the notion of iterated commutators
\begin{eqnarray}
R_{0}\equiv R_{0}(S):= S,\quad  R_{1}\equiv R_{1}(S) :=[T,S],\;\quad
\dots,\quad \; R_{l} \equiv R_{l}(S):= [T,R_{l-1}(S)].
\label{R1}
\end{eqnarray}
one has for even $n=2l,\quad  l=0,1,2,3,\dots,$
\begin{equation}
F_{2l}(S;S)=\beta^{2l}(R_{l};R_{l})_{T}=
\beta^{2l-1}\langle[R_{l}^+R_{l-1}-R_{l-1}R_{l}^+]\rangle_{T},
\label{FRR1}
\end{equation}
where $R_l^+$ denotes the Hermitian conjugate of $R_l$ and by definition
$R_{-1}\equiv X_{ST}$ is a solution of the operator equation
\begin{equation}
S = [T,X_{ST}]
\label{oe}.
\end{equation}
In view of relations (\ref{FRR1}), the functional $F_{2l}(S;S)$  may be called "Bogoliubov-Duhamel inner product of order $l$".

 In the case of odd $ n=2l+1,\quad  l=0,1,2,3,\dots$, one has
 \begin{equation}
F_{2l+1}(S;S)=
\beta^{2l}\langle[R_{l}^+R_{l}+R_{l}R_{l}^+]\rangle_{T}.
\label{FRRa}
\end{equation}

It is remarkable that both formulas (\ref{FRR1}) and (\ref{FRRa}) may be written as one formula in the form
\begin{eqnarray}
 F_{n}(S;S)=\beta^{n-1}[(-1)^{n-1}\langle R_{n-1}S^+\rangle_T + \langle R_{n-1}^+S\rangle_T],\quad
 n=0,1,2,...
\label{bfa}
\end{eqnarray}
which is more useful in some cases. Eq.(\ref{bfa}) is obtained from Eq.(\ref{FRR1}) and Eq.(\ref{FRRa}) with the successively using  of the following identities
\begin{eqnarray}
 \langle R_{l}^+R_{l-1}\rangle_T=\langle R^+_{l-1}R_{l}\rangle_T,
\quad \langle R_{l-1}R_{l}^+\rangle_T=\langle R_{l}R_{l-1}^+\rangle_T, \nonumber\\
\langle R_l^+R_{l}\rangle_T=\langle R_{l-1}^+R_{l+1}\rangle_T,\quad  \langle R_lR_{l}^+\rangle_T=\langle R_{l+1}R_{l-1}^+\rangle_T,  \nonumber\\
\end{eqnarray}
which are simple consequence of the cyclic property of the trace and the definition of the iterated commutators (\ref{R1}). Setting $n=0$ in Eq. (\ref{bfa})
 one gets the relation \begin{equation}
F_0(S,S)=-\beta^{-1}[\langle R_{-1}S^+\rangle_T - \langle R^+_{-1}S\rangle_T].
\label{bdip}
\end{equation}
It presents the Bogoliubov- Duhamel inner product as a thermodynamic mean value of the solution of Eq.(\ref{oe}) and $S$.

From the other side  the functionals $F_{n}(S;S)$ defined by Eq.(\ref{BT10ab}) may be obtained using the definition of the moments of the DSF Eq.(\ref{dm1})
 and the relation (\ref{dbp}), i.e.
\begin{equation}
F_{n}(S;S)=\beta^{n-1}[M_{n-1}(S) +  M_{n-1}(S^+)],\quad n=0,1,2,...
\label{sr}
\end{equation}
Now, it is easy to check that
\begin{eqnarray}
M_{n-1}(S) =(-1)^{n-1} \langle R_{n-1}S^+ \rangle_T, \quad
M_{n-1}(S^+)=\langle R_{n-1}^+S\rangle_T,
\label{sumr}
\end{eqnarray}
which are not but  the well known sum rules for the moments of the DSF in the linear response theory (see, e.g.\cite{F80} and \cite{PS04}, where some sum rules for the lowest moments are presented).
The expressions Eq.(\ref{sumr})  provide an algebraic way to evaluate the moments of the DSF.

The expressions (\ref{FRR1}) and  (\ref{FRRa}) take a simpler  form  in the case  $S=S^+$ that is required for the calculation in the text of the paper.
From Eq. (\ref{bfa})  with the observation that $R_n^+=(-1)^nR_n,\quad n=0,1,2,...$ we obtain
\begin{eqnarray}
F_{n}(S;S)=2(-1)^{n+1}\beta^{n-1}\langle R_{n-1}S\rangle_T,\quad
 n=0,1,2,...
\label{nnew}
\end{eqnarray}
or alternatively from Eq.(\ref{sr})

\begin{eqnarray}
F_{n}(S;S)=2\beta^{n-1}M_{n-1}(S),\quad n=0,1,2,...
\label{VRNT}
\end{eqnarray}

We note that $K_{SS}(\lambda)$ has the properties of the scalar product and
has been studied separately \cite{S00,R09} because of its relation with the Wigner, Yanase and  Dyson (WYD) skew information
\begin{eqnarray}
I_{S,S}(\lambda)=-\frac{1}{2}[Z({T})]^{-1}Tr\left([e^{-\beta\lambda T},S^+][e^{-\beta(1-\lambda)T},S]\right),\quad
0 \leq \lambda \leq 1.
\label{wyd1}
\end{eqnarray}
 via the relation
\begin{equation}
K_{SS}(\lambda)=K_{SS}(0)-I_{SS}(\lambda).
\label{wid1}
\end{equation}
The Eq. (\ref{wid1})  allows to emphasize the relation of $F_{n}(S;S)$ with the  WYD information (\ref{wyd1}). In our further consideration we shall follow ref. \cite{R09}.
Using the derivatives
\begin{equation}
\frac{d}{d\lambda}e^{-\beta \lambda T} = -\beta e^{-\beta \lambda T}T, \qquad \frac{d}{d\lambda} e^{-\beta(1-\lambda) T} =\beta e^{-\beta(1-\lambda) T}T.
\label{2p}
\end{equation}
after some simple algebra one obtains
\begin{equation}
\frac{d^n}{d\lambda^n}K_{SS}(\lambda)=(-1)^n\beta^nK_{SR_n}(\lambda),\quad n=0,1,2,...,
\label{sp2}
\end{equation}
where $ R_n=[T,R_{n-1}]$, and $R_{0}=S$.
The  properties of these derivatives are studied in details in \cite{R09} emphasizing the relation with the
Bogoliubov and Tyablicov Green's function method. This seems to provide an  useful method for their calculation.
Our finding is that  (if $S^+=S$)
\begin{equation}
F_{n}(S;S)=2\beta^{-1}\frac{d^{(n-1)}}{d\lambda^{(n-1)}}K_{SS}(\lambda)|_{\lambda=0},\qquad n=1,2,...
\label{nnew1}
\end{equation}
or equivalently
\begin{equation}
F_{n}(S;S)=-2\beta^{-1}\frac{d^{(n-1)}}{d\lambda^{(n-1)}}I_{SS}(\lambda)|_{\lambda=0},\qquad n=1,2,...,
\label{nnew2}
\end{equation}
due to Eq.(\ref{wid1}).
Differentiating and integrating  Eq.(\ref{Bsr}) with respect to $\lambda$, one obtains  \cite{R09}:
\begin{equation}
  \frac{d^n}{d\lambda^n}K_{SS}(\lambda)=\beta^n \int_{-\infty}^{\infty}d\omega  \omega^n Q_{S}(\omega;\lambda),\quad n=0,1,2,...
\label{f1}
\end{equation}
and
\begin{equation}
\int d\lambda K_{SS}(\lambda)=\beta^{-1} \int_{-\infty}^{\infty}d\omega  \omega^{-1} Q_{S}(\omega;\lambda),
\end{equation}
respectively.
 At $\lambda=0$, these are proportional to the moments of the DSF
\begin{equation}
M_n(S):=\int_{-\infty}^{\infty}d\omega  \omega^n Q_{S}(\omega),\quad n=-1,0,1,2,...
\end{equation}
 Finally, let us note that from Eqs.(\ref{nnew1}) and (\ref{f1}) we obtain Eq.(\ref{VRNT}) as it must be.

\end{document}